\documentclass[aps,prl,reprint]{revtex4-1}\usepackage{blindtext}
\usepackage{graphicx}
\usepackage{amsmath}
\usepackage{amssymb}
\usepackage[english]{babel}
\usepackage{xcolor}
\usepackage{siunitx}
\usepackage[colorlinks=true, pdfstartview=FitV, linkcolor=blue,
citecolor=blue, urlcolor=blue]{hyperref}
\usepackage[capitalise]{cleveref}
\usepackage{float}

\begin{document}

\title{Squeezed hole spin qubits in Ge quantum dots with ultrafast gates at low power}

\author{Stefano Bosco}
\author{M\'{o}nica Benito}
\author{Christoph Adelsberger}
\author{Daniel Loss}

\affiliation{Department of Physics, University of Basel, Klingelbergstrasse 82, 4056 Basel, Switzerland}

\begin{abstract}
Hole spin qubits in planar Ge heterostructures are one of the frontrunner platforms for scalable quantum computers. 
In these systems, the spin-orbit interactions permit efficient all-electric qubit control.
We propose a minimal design modification of planar devices that enhances these interactions by orders of magnitude and enables low power ultrafast qubit operations in the GHz range.
Our approach is based on an asymmetric potential that strongly squeezes the quantum dot in one direction.
This confinement-induced spin-orbit interaction does not rely on microscopic details of the device such as growth direction or strain, and could be turned on and off on demand in state-of-the-art qubits.
\end{abstract}

\maketitle

\paragraph{Introduction.}

Holes in germanium (Ge) are promising candidates for semiconductor based quantum information processing~\cite{scappucci2020germanium}.
Ge is one of the frontrunner materials for spin qubits because the noise caused by hyperfine interactions can be strongly suppressed by isotopic purification \cite{itoh1993high}, and holes do not suffer from valley degeneracies, a limiting factor for electrons \cite{vandersypen2019semiconductor}.
In addition, holes exhibit a strong spin-orbit interaction (SOI), which enables electrically controlled single \cite{watzinger2018germanium,froning2020ultrafast,wang2020ultrafast,Hendrickxsingleholespinqubit2019} and two-qubit gates~\cite{hendrickx2020fast}.
Among the several possible architectures, quantum dots in planar Ge/SiGe heterostructures are one of the most advanced. Singlet-triplet encoding~ \cite{jirovec2020singlet} as well as a four-qubit quantum processor have been  demonstrated~\cite{hendrickx2020four}, and the high degree of compatibility of these systems with  CMOS technology paves a  way towards scalable quantum computers~\cite{veldhorst2017silicon}.

In current Ge/SiGe devices, the quantum dots are rather symmetric, with a lateral confinement much smoother than the heterostructure width.
This design results in a SOI that is cubic in momentum~\cite{terrazos2018theory} and that inherently relies on the small anisotropies of the valence band of Ge \cite{wang2019suppressing}. 
In contrast, in Ge wires the holes are tightly confined in two directions and show a much larger direct Rashba (DR) SOI~\cite{DRkloeffel1,DRkloeffel2,froning2020strong,DRkloeffel3,doi:10.1002/adma.201906523,vukus2018single} that is linear in momentum and only weakly depends on valence band anisotropies.
The DRSOI is consistent with the faster Rabi oscillations observed in wires~\cite{froning2020ultrafast,wang2020ultrafast} compared to planar qubits~\cite{Hendrickxsingleholespinqubit2019}.

In this letter, we propose a minimal modification of state-of-the-art Ge/SiGe qubits that results in orders of magnitude larger Rabi frequencies, enabling  power efficient ultrafast gates.
Our approach is based on a squeezed dot, where one of the lateral directions is tightly confined. This design takes full advantage of the hole physics by recovering the large DRSOI typical of wires and in contrast to alternative proposals~\cite{PhysRevB.103.085309} only weakly depends on the growth direction of the heterostructure. The DRSOI also opens up to the possibility of strongly coupling these qubits to microwave resonators~\cite{landig2018coherent,mi2018coherent}, potentially enabling long-range interactions between distant qubits and surface code architecture~\cite{PhysRevLett.118.147701}.

\paragraph{Theoretical model.}

\begin{figure}[t]
\centering
\includegraphics[width=0.45\textwidth]{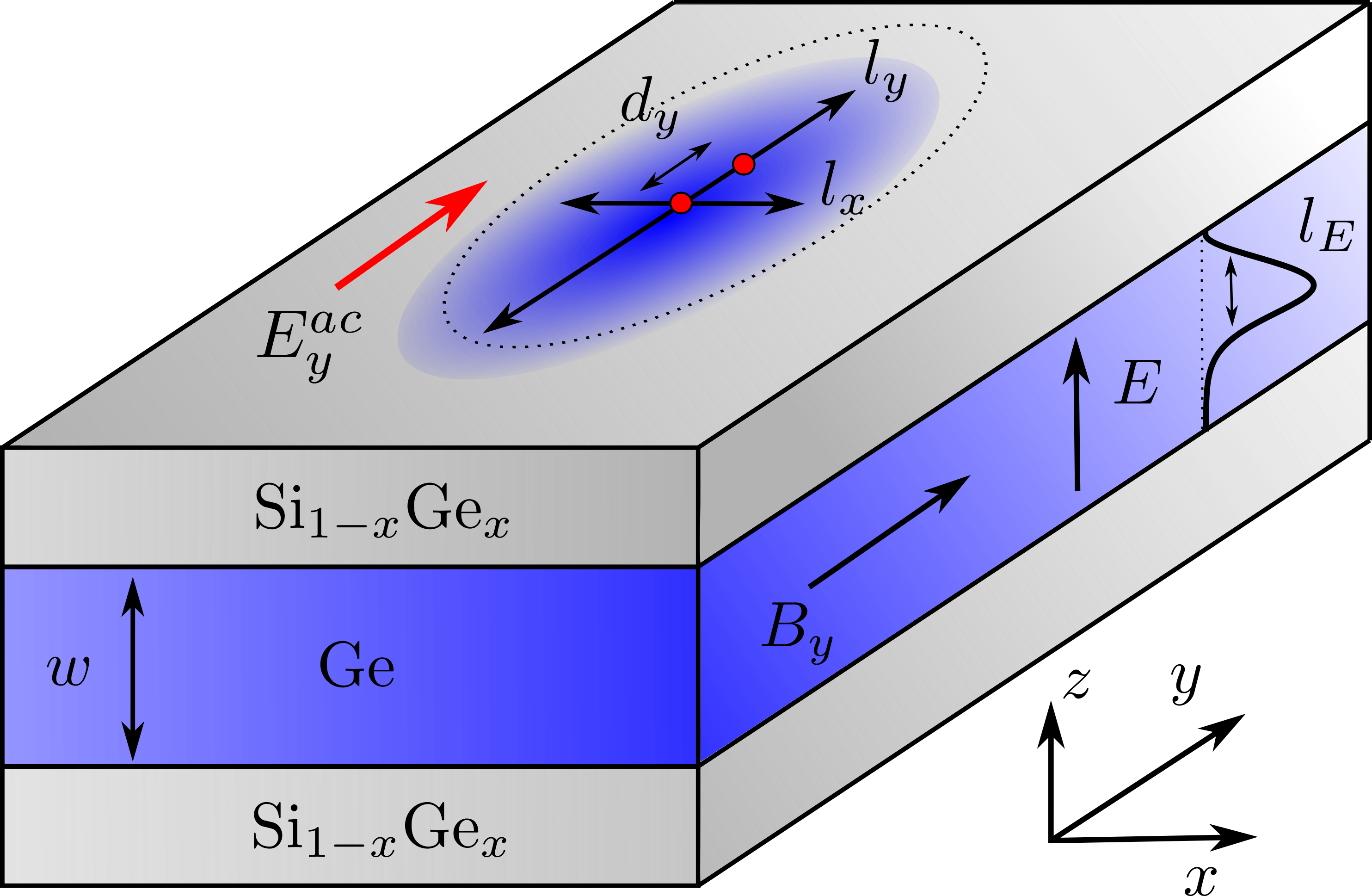}
\caption{\label{fig:dot-sketch} Squeezed hole qubit. A quantum dot is defined in a Ge well of width $w$ sandwiched between two $\text{Si}_{1-x}\text{Ge}_x$ layers. The hole wavefunction is localized within a region of width $l_E$ from the top interface of the heterostructure by a dc electric field $E$ and is confined in the $(x,y)$ plane by a anisotropic harmonic potential parameterized by the lengths $l_{x}$ and $l_y$. Different spin states are gapped by an in-plane magnetic field $B_y$. An ac electric field $E_y^{ac}$ shifts the dot time-dependently by $d_y$, resulting in ultrafast Rabi oscillations. }
\end{figure}

We examine the Ge quantum dot sketched in Fig.~\ref{fig:dot-sketch} and modelled by the Hamiltonian 
\begin{equation}
\label{eq:total-H-3D}
H=\left(\!\gamma_1+\frac{5\gamma_s}{2}\!\right)\!\frac{p^2}{2m} - \frac{\gamma_s}{m} \left( \textbf{p} \cdot \textbf{J}\right)^2+|b| \varepsilon_0 J_z^2  +V_\text{C}-eEz+H_\textbf{B}\ , 
\end{equation}
where $\textbf{p}=-i\hbar \nabla$ is the canonical momentum [$p^2=-\hbar^2\nabla^2$] and $\textbf{J}=(J_x,J_y,J_z)$ is the vector of spin 3/2 matrices. Heavy holes (HH) and light holes (LH) are mixed by the isotropic Luttinger-Kohn (LK) parameters  $\gamma_1\approx 13.35$  and $\gamma_{s}\equiv (\gamma_2+\gamma_3)/2\approx 4.96$ \cite{WinklerSpinOrbitCoupling2003}, and  by the Bir-Pikus  strain energy $b \varepsilon_0$ \cite{bir1974symmetry}, where $b=-2.16$~eV and $\varepsilon_0\equiv\varepsilon_{\parallel}-\varepsilon_{zz}\approx 1.74\varepsilon_{\parallel}$. The uniaxial strain  caused by the mismatch of the lattice constants in the heterostructure is described by the homogeneous strain tensor  $\varepsilon_{ij}\approx \delta_{ij}\varepsilon_{ii}$, with $\varepsilon_{xx}=\varepsilon_{yy}\equiv \varepsilon_{\parallel}$ and $\varepsilon_{zz}=-2C_{12}\varepsilon_\parallel/C_{11}$ \cite{terrazos2018theory, wang2019suppressing} [$C_{ij}$ are the elastic constants of Ge].  
The confinement energy $V_\text{C}=V_z(z)+\sum_{i=x,y} \hbar\omega_i r_i^2/2l_i^2$ comprises an abrupt potential $V_z$ modelling the boundaries of a heterostructure of width $w$, and an electrostatic potential in the $\textbf{r}=(x,y)$ plane, parameterized by the harmonic lengths $l_i$ and by the frequencies $\omega_i\equiv\hbar\gamma_1/ml_i^2$. 
While dc electric fields in the $\textbf{r}$ plane have no effect on the system, the externally tunable electric field $E>0$ compresses the wavefunction within a length $l_E\equiv(\hbar^2\gamma_1/2meE)^{1/3}\approx 8$~nm$\times E^{-1/3}$ from the top boundary of the heterostructure \cite{PhysRevResearch.2.043180} and controls the SOI. To simplify the notation, throughout the paper $E$ is given in V/$\mu$m.  The lengths $l_{x,y,E}$ introduced here are parameters that model the electrostatic potential and depend on an average hole mass $m/\gamma_1$ [$m$ is the electron mass] equal for HHs and LHs.
To define the qubit, we include an external magnetic field $\textbf{B}$, typically of a few hundreds of milli-Tesla.  The resulting Hamiltonian $H_\textbf{B}=H_Z+H_O$ comprises  the Zeeman energy $H_Z=2\mu_B \textbf{B}\cdot(\kappa \textbf{J}+q \textbf{J}^3)$  \cite{WinklerSpinOrbitCoupling2003} and the orbital contribution $H_O\approx-2e\gamma_s \{\textbf{A}\cdot \textbf{J},\textbf{p}\cdot \textbf{J}\}/m$ coming from the Peierls substitutions $\textbf{p}\rightarrow \pmb{\pi}=\textbf{p}+e\textbf{A}$, with $\textbf{A}=-(B_z y,0,B_y x-B_x y)$ being the vector potential. We neglect irrelevant shifts of the dot and corrections~$\mathcal{O}(\textbf{B}^2)$.  
 
\paragraph{Optimal conditions for the DRSOI.}

\begin{figure}[t]
\centering
\includegraphics[width=0.5\textwidth]{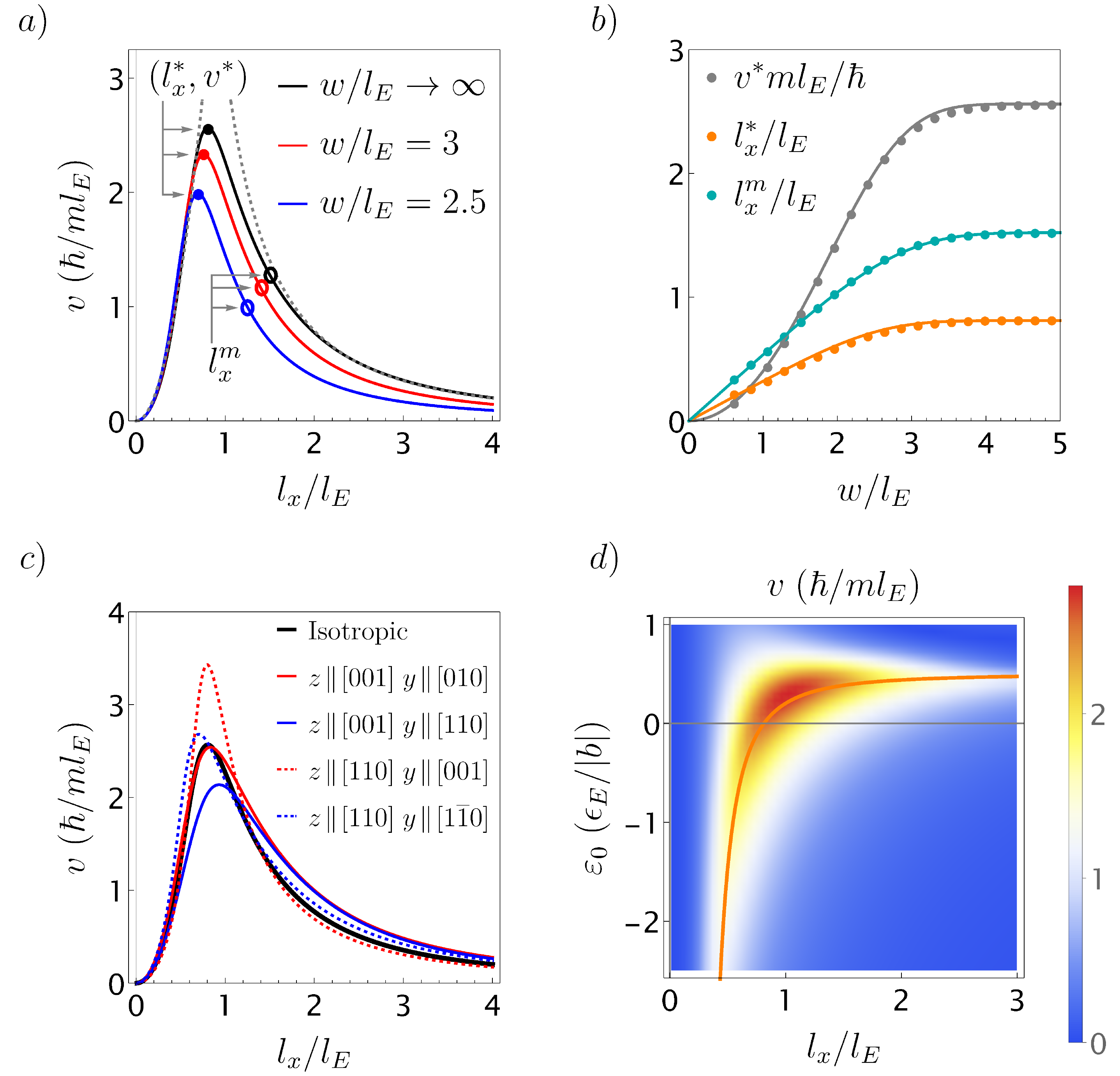}
\caption{\label{fig:v_so} DRSOI in a squeezed Ge dot. The spin-orbit velocity $v$ and the lengths are in units of $\hbar/ml_E\approx 9.53$~meV~nm$\times E^{1/3}/\hbar$ and  $l_E\approx 8$~nm$\times E^{^{-1/3}}$, respectively. $E$ is in~V/$\mu$m.
In a) we show $v$  against  $l_x$ for different values of $w$. The gray dashed lines are the asymptotic limits $v= 5.1 \hbar l_x^2  /m l_E^3$ and $v= 3.16 \hbar l_E  /m l_x^2$ obtained when $w\gg l_E$ for small and large $l_x/l_E$, respectively.   The largest SOI $v=v^*$ is reached at $l_x=l_x^*$, while $v=v^*/2$ at $l_x=l_x^m>l_x^*$. The dependence of these parameters on $w$ is shown in b); dots and lines are respectively numerical and the fitted results as discussed in the text.
In c) and d), we focus on the $w\gg l_E$ case. In c) we study the effect of the anisotropies  $\gamma_2 \neq \gamma_3$ of the LK Hamiltonian for different confinement directions, and in d) we consider strained Ge. The strain $\varepsilon_0$ is in units of $\epsilon_E/|b| \approx 0.37\%\times E^{2/3} $. The largest SOI is reached along the orange curve $|b|\varepsilon_0/\epsilon_E=0.5- 0.2/(0.18-l_x/l_E)^2$. }
\end{figure}
 
In the squeezed dot sketched in Fig.~\ref{fig:dot-sketch},  $E$ induces a DRSOI $H_\text{SO}= v p_y \sigma_x$, which tends to align the groundstate quasi-degenerate Kramers partners to the $x$-direction.
To predict the optimal design for the DRSOI, we estimate the spin-orbit velocity $v$  by  first diagonalizing $H$ at $p_y=0$ and $\textbf{B}=0$, and then projecting $H_1=-2\gamma_s p_y \{J_y,p_xJ_x+p_zJ_z\}/m$ onto the groundstate subspace~\cite{bosco2020hole}. Here, $\{A,B\}=(AB+BA)/2$.
The SOI depends on the lengths $l_{x,E}$ and $w$, and on the strain $\varepsilon_0$.

We first set $\varepsilon_0=0$, and in Fig.~\ref{fig:v_so}a), we show how $v$ varies as a function of $l_x/l_E$ for heterostructures with different widths $w$.
When $l_x\ll l_E$, the SOI is accurately described by the expansion $\hbar v\approx 5.1 \hbar^2 l_x^2  /m l_E^3= 0.76 eE l_x^2$, independent of  $w$.  
As the ratio $l_x/l_E$ increases,  $v$ reaches the maximal value $v^*$ at $l_x=l_x^*$,  and then decays as $v \propto l_E/l_x^2 \propto E^{-1/3}$ for $l_x\gg l_E$. 

The position and value of the maximal SOI depend on $w$ and these dependences are very well approximated by the fitting formulae $v^*=2.56 \text{erf}(0.14 w^2/l_E^2 )\hbar/ml_E$ and $l_x^*=0.81  l_E \sqrt{\text{erf}\left(0.14 w^2/l_E^2\right)}$, as shown in Fig.~\ref{fig:v_so}b). 
The optimal SOI saturates to $\hbar v^*=2.56\hbar^2/ml_E\approx 25$~meV~nm$\times E^{1/3}$ when $w\gtrsim 3 l_E\approx 24$~nm$\times E^{-1/3}$; this condition is easily met in state-of-the-art devices, where $w\in [15,30]$~nm and $E\gtrsim 1$~V/$\mu$m \cite{scappucci2020germanium}. We remark that the case $w\gg l_E$ also describes inversion layers.

The condition for the optimal length $l_x^*\approx 0.81 l_E$  requires a strong harmonic potential $\hbar\omega_x= 24$~meV$\times E^{2/3}$ that compresses the wavefunction in a region shorter than $6.5$~nm$\times E^{-1/3}$. While not unrealistic for industry standards \cite{1175829}, we can relax this constraint by introducing the quantity $l_x^m$, defined as the largest value of $l_x$ that guarantees $v>v^*/2$, see Fig.~\ref{fig:v_so}a).
As shown in Fig.~\ref{fig:v_so}b), an excellent fitting formula for $l_x^m$ is $ l_x^m=1.52 l_E \sqrt{\text{erf}\left(0.11 w^2/l_E^2\right)} $, resulting in the experimentally accessible length  $l_x^m\approx 12$~nm$\times E^{-1/3}$ at $w\gg l_E$.

The isotropic LK Hamiltonian in Eq.~\eqref{eq:total-H-3D} neglects small cubic anisotropies $\propto(\gamma_3-\gamma_2)/\gamma_1\approx 0.1$ \cite{WinklerSpinOrbitCoupling2003}. When these terms are included, $v$ depends on the alignment between  confinement and crystallographic axes.
As shown in Fig.~\ref{fig:v_so}c), a more refined analysis analogous to Ref.~\cite{bosco2020hole} shows that the isotropic approximation describes well the system, but there are special orientations at which the DRSOI is enhanced: the DRSOI is largest when $z\parallel [110]$ and $y\parallel [001]$ \cite{DRkloeffel3, bosco2020hole}. 
In contrast to other proposals \cite{PhysRevB.103.085309} for obtaining DRSOI in Ge heterostructures, in our approach this particular growth direction is convenient but not required. Also, because here the DRSOI originates from the confinement potential and not from the small anisotropies of Ge, the maximal SOI $\hbar v^*$ is more than 5 times larger than in~\cite{PhysRevB.103.085309} at comparable electric fields.
 
The strong DRSOI persists in strained heterostructures. 
In Fig.~\ref{fig:v_so}d), we analyze the dependence of $v$ on $l_x/l_E$ in a strained device with $w\gg l_E$. 
Here, we measure the strain $\varepsilon_0$ in units of $\epsilon_E/|b|$, with $\epsilon_E\equiv\hbar^2\gamma_1/2ml_E^2$ being the electric energy. 
Compressive strain with $\varepsilon_0<0$  tends to align the spin quantization axis to the $z$-direction, thus reducing the HH-LH mixing and the DRSOI. However, in the range of parameters studied, the maximal SOI lying on the fitted orange curve $|b|\varepsilon_0/\epsilon_E=0.5- 0.2/(0.18-l_x/l_E)^2$ is only halved. In this case, a tighter lateral confinement is required to reach the optimal DRSOI and the wavefunction needs to be further squeezed to $l_x\approx 0.44 l_E$. 
In contrast to Ge/Si core/shell wires, where the strain increases the small gap between ground and first excited states \cite{DRkloeffel1,DRkloeffel2,DRkloeffel3} and to electron-based devices, where strain removes the valley degeneracy \cite{vandersypen2019semiconductor}, in planar hole systems strain is not fundamentally required and could potentially be minimized.

\paragraph{Rabi driving in a squeezed quantum dot.}

\begin{figure}[t]
\centering
\includegraphics[width=0.5\textwidth]{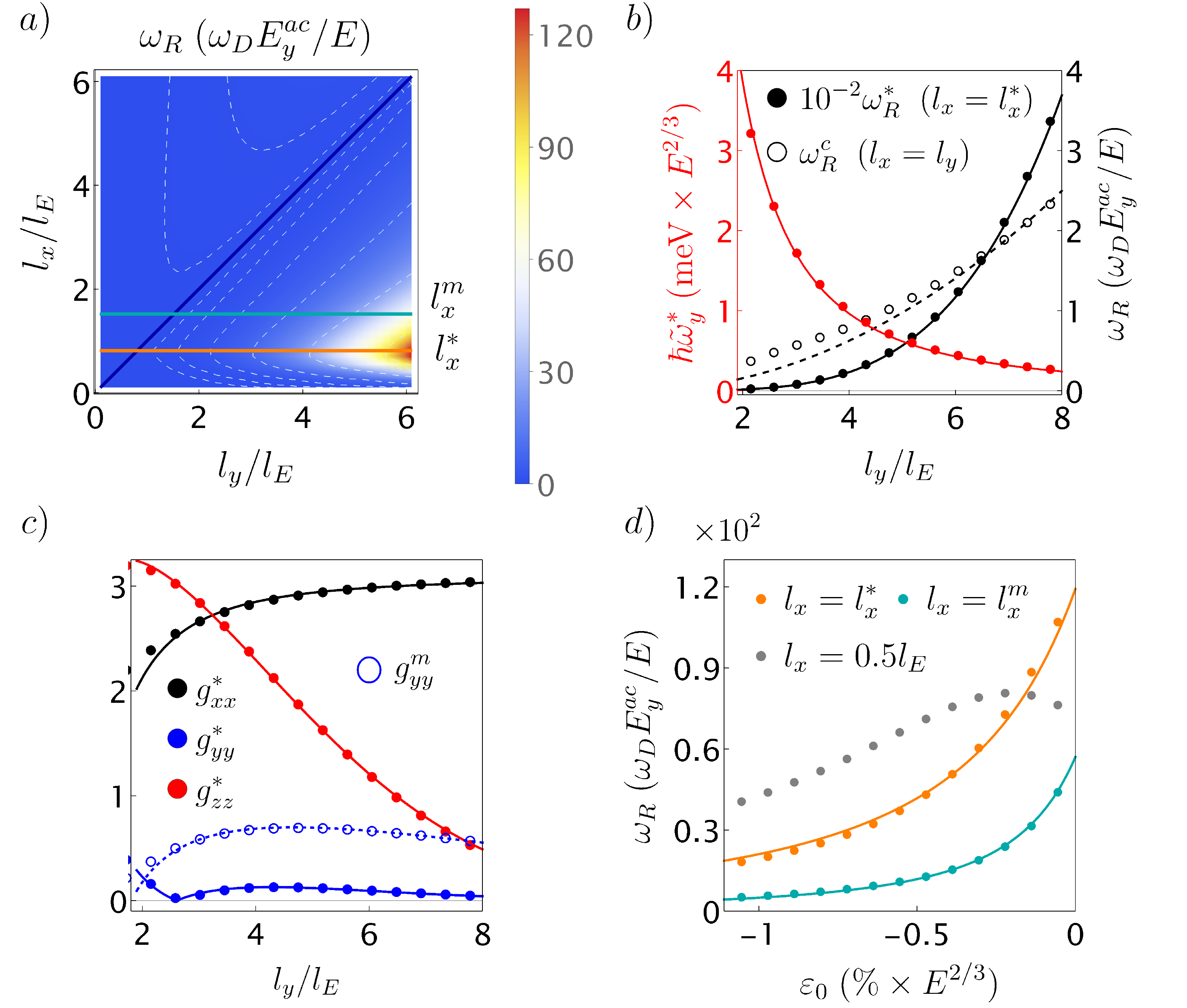}
\caption{\label{fig:dot_simulation} Rabi driving of a squeezed dot.
We compare analytical results (lines) with numerical simulations (dots) of a 3-dimensional dot driven by a field with amplitude $E_y^{ac}$ and frequency $\omega_D$. Here, $w\gg l_E$.  In a) we analyze the dependence of Rabi frequency $\omega_R$ on the aspect ratio of the dot. 
In b) we show $\omega_R$ (black) against $l_y$ in the DR ($l_x=l_x^*=0.81l_E$) and cubic  ($l_x=l_y$) SOI regime. In the latter case, we use the anisotropic LK Hamiltonian and   $z\parallel[001]$. To facilitate the comparison, $\omega_R^*$ has been reduced by a factor $10^2$. The energy gap $\tilde{\omega}_y^*$ at $l_x=l_x^*$ is shown in red.
In c) we show the $g$ tensor at $l_x=l_x^*$ against $l_y$; hollow dots show $g_{yy}$ at $l_x=l_x^m$. 
In d) we examine $\omega_R$ at $l_y=6l_E$ against $\varepsilon_0$ for different values of $l_x$; solid lines mark the fitting formulae discussed in the text.  In the left (bottom) label of b (d) $E$ is in V/$\mu$m. }
\end{figure}

In a squeezed Ge quantum dot, the large DRSOI enables ultrafast qubit operations.
In fact, a time-dependent shift of the dot caused by ac in-plane fields $E_{x,y}(t)$ can drive transitions between different qubit states via EDSR \cite{PhysRevB.74.165319}.
This effect can be understood by moving to a frame that oscillates with the center of the dot at position $\textbf{r}_i=\textbf{d}_i(t)\equiv eE_i(t)l_i^2/\hbar\omega_i$ via the time-dependent translation $T=e^{-i\textbf{p}\cdot \textbf{d}/\hbar}$. In this frame, the hole still evolves according to $H$, but feels additionally the external drive $H_D(t)=-i\hbar T^\dagger\partial_t T=-\textbf{p}\cdot \partial_t\textbf{d}(t)$.
When the dot is strongly confined in the $x$-direction, the oscillation is restricted to the $y$-direction, i.e. $H_D(t)\approx -p_y \partial_td_y(t)$. This system is  modelled by the wire Hamiltonian
\begin{equation}
H_\text{W}= \frac{p_y^2}{2 \tilde{m}}+\frac{\tilde{m}\tilde{\omega}_y^2}{2}y^2+vp_y\sigma_x+\frac{\mu_B}{2}\ \textbf{B}\cdot \tilde{\underline{g}}\cdot \pmb{\sigma}-p_y\partial_td_y(t) \ ,
\end{equation}
acting on the Kramer partners $|\uparrow\rangle$ and $|\downarrow\rangle$.
Here, we introduce a matrix   $ \tilde{g}_{ij}=\delta_{ij}(\alpha_i-\beta_i p_y^2/\hbar^2)$ of wire $g$-factors, which is diagonal because of symmetry \cite{PhysRevResearch.2.033036, bosco2020hole} and includes momentum dependent corrections $\beta_i$ \cite{DRkloeffel2,DRkloeffel1}. The orbital gap $\tilde{\omega}_y$ differs from the frequency $\omega_y$ because of the effective  mass $\tilde{m}$, i.e. $\tilde{\omega}_y=\omega_{y}\sqrt{m/\gamma_1 \tilde{m}}$. In analogy, the dot width is $\tilde{l}_y=l_y\sqrt[4]{m/\gamma_1 \tilde{m}}$. 

When the drive and the Zeeman energy are much smaller than $\tilde{\omega}_y$, an effective quantum dot theory is obtained by projecting $H_\text{W}$ onto the groundstates $\Psi_{\uparrow\downarrow} = \psi(y) e^{-i\sigma_x y/l_{so}}|\uparrow\downarrow\rangle$  of $H_\text{W}$ at $\textbf{B}=E_y(t)=0$.
The transformation $e^{-i\sigma_x y/l_{so}}$ exactly removes the SOI and we introduce the spin-orbit length $l_{so}\equiv \hbar/m^*v$; $\psi(y)=e^{-y^2/2\tilde{l}_y^2}/\sqrt[4]{\pi \tilde{l}_y^2}$.
The resulting  qubit Hamiltonian  is
\begin{equation}
H_\text{Q}=\frac{\mu_B}{2}\textbf{B}\cdot \underline{g}\cdot \pmb{\sigma}+ \epsilon_D(t) \sigma_x \ ,
\end{equation}
where the dot $g$-factors \cite{DRkloeffel2,DRkloeffel1,froning2020strong} and the driving are
\begin{subequations}
\label{eq:Rabi-drive-and-g}
\begin{align}
\label{eq:g-fact}
g_{xx}&=\alpha_x-\frac{\beta_x}{2\tilde{l}_y^2} \ \ \text{and} \ \ g_{ii}=\!\left(\!\alpha_i-\frac{\beta_i}{2\tilde{l}_y^2\!}\right)e^{-\frac{\tilde{l}^2_y}{l_{so}^2}}  \ , \\
\label{eq:Rabi-drive}
\epsilon_D(t)&= \frac{\partial_t d_y(t)}{l_{so}} = \frac{\tilde{l}_y}{l_{so}} \frac{e\partial_t E_y(t)\tilde{l}_y}{\hbar\tilde{\omega}_y} \ ,
\end{align}
\end{subequations}
with $i={y,z}$.
If we now specialize our analysis to the case $\textbf{B}=B_y \textbf{e}_y$ and consider a harmonic drive $E_y(t)=E_y^{ac}\sin(\omega_D t)$, at the resonance $\omega_D=g_{yy}\mu_B B_y$ the qubit shows Rabi oscillations with frequency 
\begin{equation}
\label{eq:Rabi-freq}
\omega_R= \frac{l_y}{2l_{so}}\left(\frac{l_y}{l_E}\right)^3\frac{E_y^{ac}}{E}\omega_D \ .
\end{equation}
Because our analysis treats the SOI exactly, Eqs.~\eqref{eq:Rabi-drive} and~\eqref{eq:Rabi-freq} are applicable for arbitrary SOI strengths. At resonance, they also agree with the perturbative results in Ref.~\cite{PhysRevB.74.165319} and correctly vanish when $E_y$ is static. 
By substituting the operators $T$ with magnetic translations~\cite{froning2020strong}, one can show that at resonance the orbital effects only give corrections $\mathcal{O}(\textbf{B}^2)$ and are  neglected here.

In Fig.~\ref{fig:dot_simulation}, we use these analytical formulae to interpret the results of a full 3-dimensional numerical simulation of the dot in Fig.~\ref{fig:dot-sketch} when $w\gg l_E$. In this simulation, $\epsilon_D$ and $g_{ii}$ are obtained by discretizing the Hamiltonian in Eq.~\eqref{eq:total-H-3D} at $E_y(t)=\textbf{B}=0$, and by projecting $H_D(t)$ and $H_\textbf{B}$ onto the groundstate subspace. The energy gap between this subspace and the first excited state is the gap $\tilde{\omega}_y$.
In Fig.~\ref{fig:dot_simulation}a), we show the Rabi frequency $\omega_R$  produced by $E_y(t)$ as a function of the aspect ratio of the dot. We first neglect the strain and set $\varepsilon_0=0$. 
When $l_x=l_y$ (blue line) the dot is isotropic and $\omega_R$ vanishes. In contrast, $\omega_R$ is strongly enhanced at $l_x\sim l_E$ and $l_y\gtrsim 2 l_E$, where the DRSOI is large.

In Fig.~\ref{fig:dot_simulation}b), we examine in more detail these two cases. When $l_x=l_y$, a Rabi frequency $\omega_R^c\approx 0.039\omega_D E_y^{ac} l_y^2/E l_E^2$  consistent with a cubic SOI~\cite{terrazos2018theory, wang2019suppressing} is recovered for $z\parallel [001]$ when including the LK anisotropies $\gamma_2\neq \gamma_3$~\cite{wang2019suppressing}.
Because $(\gamma_2- \gamma_3)/\gamma_1\approx 0.1$, this contribution is much smaller than the DRSOI:  in the range of parameter analyzed here and for a driving field with $\omega_D=3$~GHz and $E_y^{ac}/E=2\%$ \cite{froning2020ultrafast},  we estimate a maximal value $\omega_R^c\approx 150$~MHz, in reasonable agreement with both theory~\cite{terrazos2018theory,wang2019suppressing} and experiments~\cite{Hendrickxsingleholespinqubit2019,hendrickx2020fast}. 
In contrast, in the DR regime, the Rabi frequency grows as $\omega_R\propto l_y^4$, see Eq.~\eqref{eq:Rabi-freq}, roughly independently of the growth direction, resulting in $\omega_R^*\sim 200 \omega_D E_y^{ac}/E\sim 12$~GHz at $l_x= l_x^*$. This frequency is two orders of magnitude larger than $\omega_R^c$, thus enabling faster gates at lower power.
We extract  $\tilde{m}$ and $l_{so}^*$ from the slope of $\tilde{\omega}_y^*\propto l_y^{-2}$ and $\omega_R^*$. In the regime of parameters studied  [also including strain] $\tilde{m}$ varies at most of $\pm 20\%$ from $m/\gamma_1$, resulting in a maximal variation of $\tilde{\omega}_y$ and $\tilde{l}_y$ of $\pm 10\%$ and $\pm 5\%$ from  $\omega_y$ and  $l_y$, respectively. Consequently, $l_{so}^*\approx \gamma_1 l_E / 2.56\approx 42$~nm$\times E^{-1/3}$, in good agreement with the fitted value $l_{so}^*=44$~nm$\times E^{-1/3}$. 

The $g$-factors are also described well by  Eq.~\eqref{eq:g-fact} as shown in Fig.~\ref{fig:dot_simulation}c). Assuming $\tilde{l}_y\approx l_y$, we find a good fit for  $(\alpha_{x}^*,\beta_x^*/l_E^2) =(3.09,7.72)$,  $(\alpha_{z}^*,\beta_z^*/l_E^2) =(3.92,2)$,  $(\alpha_{y}^*,\beta_y^*/l_E^2) =(0.37,5.02)$ and $(\alpha_{y}^m,\beta_y^m/l_E^2) =(0.98,6.38)$. 
When $\textbf{B}=B_y \textbf{e}_y$, at the confinement potential maximizing the DRSOI the Zeeman gap is small because $g_{yy}^*\approx 0.1$. Larger gaps are obtained  by rotating the magnetic field to the $z$-direction or by widening the dot: at $l_x=l_x^m$,  $g_{yy}^m\approx 7 g_{yy}^*$ and the DRSOI is only halved, thus still enabling above GHz Rabi oscillations. 

In Fig.~\ref{fig:dot_simulation}d), we analyze the dependence of $\omega_R$ on $\varepsilon_0$ in a strained dot with $l_y=6 l_E\approx 48$~nm$\times E^{-1/3}$. When $l_x/l_E\gtrsim l_x^*$, we find that the effect of strain is accurately captured by the fitting formula $\omega_R(\varepsilon_0)=\omega_R(0)/(1-\varepsilon_0/\overline{\varepsilon})^2$, where $\overline{\varepsilon}$ is a positive parameter that  decreases from $\overline{\varepsilon}^*=0.72 \% \times E^{2/3}$  to $\overline{\varepsilon}^m=0.42 \% \times E^{2/3}$ when $l_x=[l_x^*,l_x^m]$.
In a strained device, $\omega_R$ is enhanced by reducing $l_x$ [gray dots in the figure] or increasing $l_y$ [$\omega_R(0)\propto l_y^4$].

\paragraph{Squeezed qubit in state-of-the-art devices.}

\begin{figure}[t]
\centering
\includegraphics[width=0.5\textwidth]{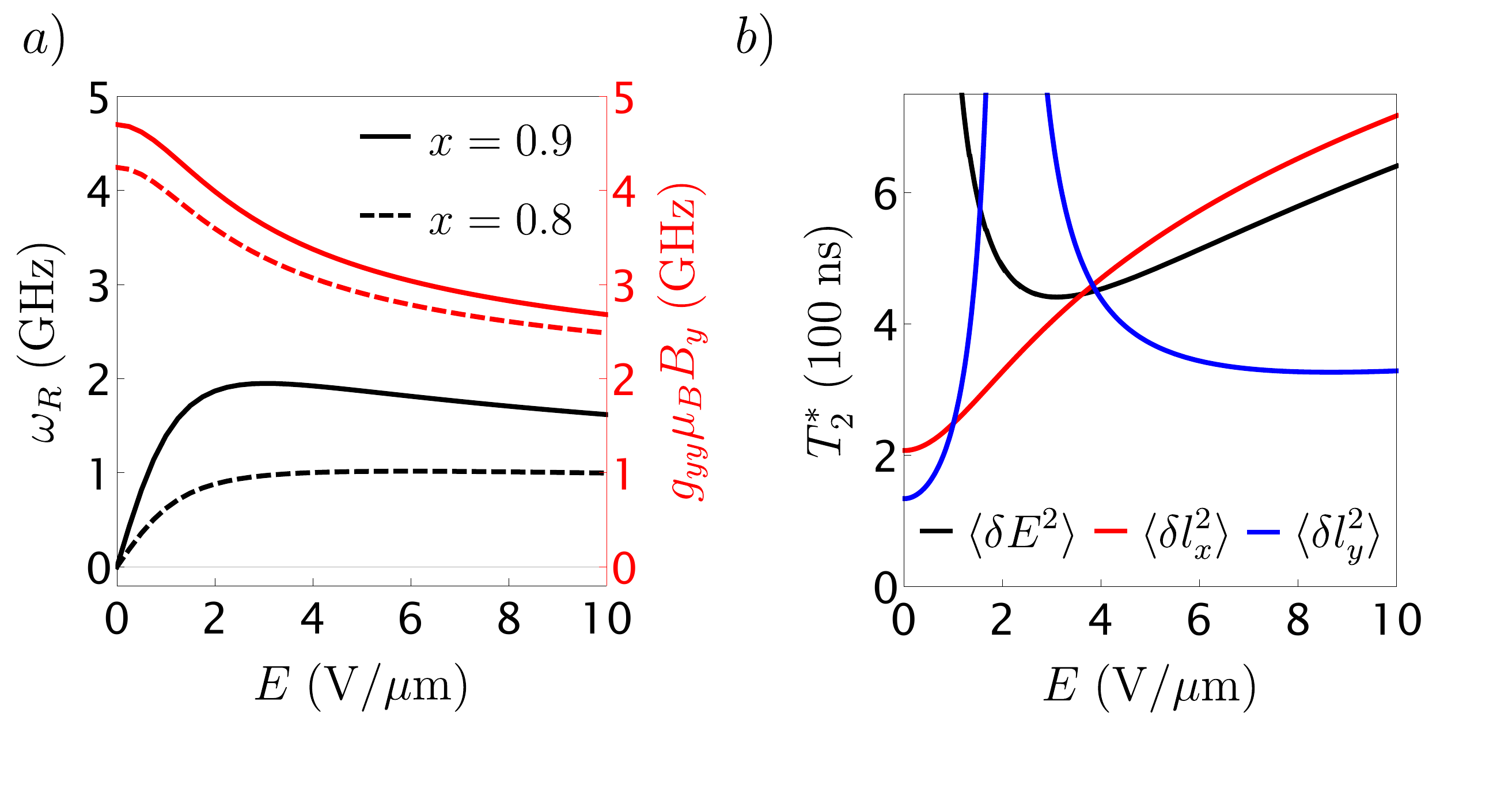}
\caption{\label{fig:real_device} Squeezed qubits in state-of-the-art devices. We simulate the dot in Fig.~\ref{fig:dot-sketch} with realistic parameters $w=20$~nm,  $l_x=10$~nm and $l_y=50$~nm; we use $B_y=0.5$~T and $E_y^{ac}=0.02$~V/$\mu$m. In a) we show the Rabi (black) and Zeeman (red) frequencies $\omega_R$ and $g_{yy}\mu_B B_y/h$ against $E$ for two devices with different concentration of Ge $x=0.9$ (solid) and $x=0.8$ (dashed) [$\varepsilon_0=-0.54\%$ and $\varepsilon_0=-1.08\%$, respectively]. In b) we estimate the dephasing at $x=0.9$; we use $\sqrt{\langle\delta E^2\rangle}/E= 10^{-3}$ and $\sqrt{\langle \delta \omega_i^2\rangle}=2\omega_i\sqrt{\langle \delta l_i^2\rangle}/l_i= 5$~$\mu$eV. }
\end{figure}

To conclude our analysis, we simulate explicitly a squeezed qubit in currently available devices \cite{strain1}.
In Fig.~\ref{fig:real_device}a), we show the Rabi and Zeeman frequencies as a function of $E$ in a dot with lateral sizes $l_x=10$~nm and $l_y=50$~nm and well width $w=20$~nm. The dot is driven at resonance by a realistic ac electric field $E_y^{ac}=0.02$~V/$\mu$m~\cite{froning2020ultrafast} and is subjected to a magnetic field $B_y=0.5$~T. 
We relate the strain $\varepsilon_0\approx 1.74\varepsilon_\parallel\approx -5.4\% (1-x)$ to the Ge concentration $x$ in the barriers [see Fig.~\ref{fig:dot-sketch}], by using the linear interpolation  $\varepsilon_{\parallel}\approx -0.62\% (1-x)/(1-0.8)$, based on the measured value $\varepsilon_{\parallel}\approx -0.62 \%$ at $x=0.8$~\cite{strain1}.
In this design,  the Zeeman energy is around $3$~GHz and is 22 to 43 times smaller than $\tilde{\omega}_y\gtrsim 420$~$\mu$eV.
The Rabi frequency is in the GHz range and doubles when $x$ changes from $0.8$ to $0.9$. These values are comparable to the estimated values in Ge nanowires \cite{DRkloeffel2} and result in ultrafast qubit gates. At the same time, because $\omega_R\propto E_y^{ac}$, the strong DRSOI enables power efficient operations and currently achieved Rabi frequencies $\omega_R\sim 100$~MHz~\cite{hendrickx2020fast,hendrickx2020four,jirovec2020singlet,Hendrickxsingleholespinqubit2019} are reached at the modest driving amplitude $E_y^{ac}\approx 2\times 10^{-3}$~V/$\mu$m.

Finally, in Fig.~\ref{fig:real_device}b), we estimate the lifetime of this qubit when left idle. Assuming 1/f charge noise, the fluctuations of $g_{yy}$ as a function of  $l_{x,y}$ and $E$ result in a dephasing time $T_2^*\approx \left[\mu_B B_y  \sqrt{\langle\delta \eta^2\rangle}  \partial g_{yy}/\sqrt{2\pi}\hbar\partial \eta\right]^{-1}\sim 300$~ns.
Here, $\eta={l_x,l_y,E}$ and we neglect logarithmic corrections of $T_2^*$~\cite{MAKHLIN2004315}.
The coherence can be improved by dynamical decoupling. Alternatively,  because the lateral confinement is controlled by tunable potentials, we envision protocols where  qubits could be squeezed on-demand only when operational, thus enabling ultrafast operations, while minimizing charge noise in the unsqueezed state.

In summary, the proposed slight modification of current planar devices based on an asymmetric confinement will push this quantum dot architecture towards new speed and coherence standards.
Our analysis is restricted to Ge but we expect similar approaches to strongly enhance the SOI in other semiconductors, such as Si, thus opening up to new possible ways to implement low power ultrafast spin qubits in planar quantum processors.

\begin{acknowledgments}
We  thank M. Russ, N. Hendrickx, and M. Veldhorst for useful discussions and for valuable comments on the manuscript.
This work was supported by the Swiss National Science Foundation and NCCR SPIN.
\end{acknowledgments}

\bibliography{references}

\end{document}